\documentstyle[aps,prl,preprint,floats,epsfig]{revtex}  

\newcommand{\dstor}{D^{(*)0}}

\newcommand{\bbar}{\overline{B}{}^0}
\newcommand{\mbc}{M_{bc}}

\begin{document}

\preprint{\tighten\vbox{\hbox{\hfil BELLE-CONF-0107}
}}


%
%

\title{
 \quad\\[1cm] \Large
Observation of Color-Suppressed $\overline{B}{}^0 \to D^{(*)0}X^0$ Decays
}

\author{The Belle Collaboration}

\maketitle

\tighten

%
%

\begin{abstract}
We report the first observation of color-suppressed 
$\bbar \to D^0 \pi^0$ and $D^{(*)0} \omega$ decays and 
evidence of $\bbar \to D^{*0} \pi^0$ and 
$D^{(*)0} \eta$.  The branching fractions are found to be 
${\cal B} (\bbar \to D^0 \pi^0) = 
(2.9\;^{+0.4}_{-0.3} \pm 0.6) \times 10^{-4}$, 
${\cal B} (\bbar \to D^0 \omega) = 
(1.7\;^{+0.5\; +0.3}_{-0.4\; -0.4}) \times 10^{-4}$, and 
${\cal B} (\bbar \to D^{*0} \omega) = 
(3.4\;^{+1.3}_{-1.1}\pm 0.8) \times 10^{-4}$.
The analysis is based on a data sample of 21.3 fb$^{-1}$ collected 
at the $\Upsilon(4S)$ resonance by the Belle 
detector at the KEKB $e^{+} e^{-}$ collider. 
\end{abstract}
%
%
%

\begin{figure}[h]
\begin{center}
\epsfig{file=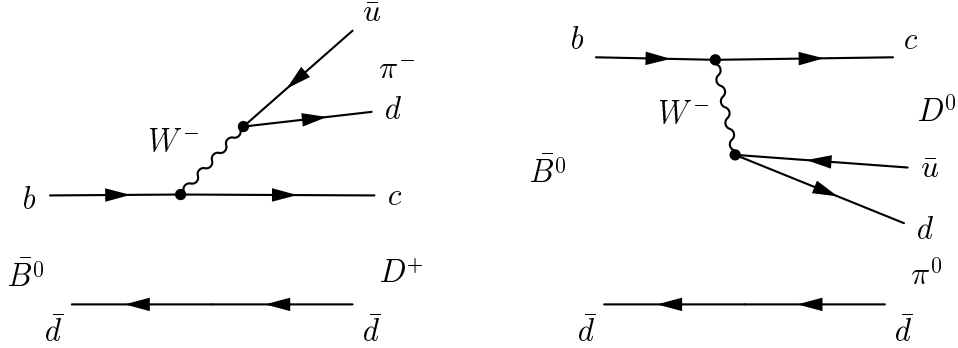,width=5in}
\medskip
\caption{The external(left) and internal(right) spectator 
diagrams for $\overline{B}\to D\pi$ decays.}
\label{dpi-feynman}
\end{center}
\end{figure}

{\renewcommand{\thefootnote}{\fnsymbol{footnote}}

\newpage

\begin{center}
  K.~Abe$^{9}$,               
  K.~Abe$^{37}$,              
  R.~Abe$^{27}$,              
  I.~Adachi$^{9}$,            
  Byoung~Sup~Ahn$^{15}$,      
  H.~Aihara$^{39}$,           
  M.~Akatsu$^{20}$,           
  K.~Asai$^{21}$,             
  M.~Asai$^{10}$,             
  Y.~Asano$^{44}$
% ...
\end{center}

\small
\begin{center}
$^{1}${Aomori University, Aomori}\\
% ...
$^{9}${High Energy Accelerator Research Organization (KEK), Tsukuba}\\
$^{10}${Hiroshima Institute of Technology, Hiroshima}\\
% ...
$^{15}${Korea University, Seoul}\\
% ...
$^{20}${Nagoya University, Nagoya}\\
$^{21}${Nara Women's University, Nara}\\
% ...
$^{27}${Niigata University, Niigata}\\
% ...
$^{37}${Tohoku Gakuin University, Tagajo}\\
% ...
$^{39}${University of Tokyo, Tokyo}\\
% ...
$^{44}${University of Tsukuba, Tsukuba}\\
% ...
\end{center}

\normalsize

\normalsize

\setcounter{footnote}{0}
\newpage

\normalsize

%
%
%


%
%

The decay modes $\overline{B}{}^0\to \dstor X^0$, where $X^0$ is a
light neutral meson, proceed via an internal
spectator diagram and are expected to be suppressed relative to the
external diagram, since the color of the
$\overline{u}$ antiquark produced by the weak current
must complement the color of the $c$ quark as shown in
Fig.~\ref{dpi-feynman}.
Studies of such color-suppressed decay modes can be used to test
models of hadronic $B$ meson decays and provide information on
final-state interactions.  
Results for color-suppressed $\overline{B}{}^0\to D^{(*)0}X^0$ decays
have been published by the CLEO collaboration~\cite{cleo-prd57}; however,
only upper limits were obtained.

In this paper, we report on a search for the color-suppressed
$\bbar\to D^0X^0$, and $D^{*0}X^0$ decay processes, where
the neutral meson $X^0$ 
is either a $\pi^0$, $\eta$, or $\omega$.
Charge conjugate modes are implicitly included in this paper.
The data sample used in this analysis was
collected with the Belle detector\cite{NIM} at KEKB\cite{kekb}.  KEKB 
is a double storage ring with 8 GeV electrons and 3.5 GeV positrons 
colliding at a 22 mrad crossing angle.
The data sample corresponds to an integrated luminosity of 
$21.3$ fb$^{-1}$ at the $\Upsilon(4S)$ resonance
and contains 22.8 million $B\overline{B}$ pairs.

%
%

Belle is a general-purpose detector containing a 1.5 T superconducting
solenoid magnet.  Charged
particle tracking covering 92\% of the total
center-of-mass (CM) solid
angle is provided by a Silicon Vertex Detector (SVD) consisting of
three concentric layers of double sided silicon strip
detectors, and a 50-layer Central Drift Chamber
(CDC). Particle identification is accomplished by combining
the responses from an array of Silica Aerogel \v Cerenkov Counters (ACC)
and a 
Time of Flight Counter system (TOF) with {\it dE/dx}
measurements in the CDC. The combined response of the three systems
provides at least 2.5$\sigma$ $K/\pi$ separation for laboratory
momentum up to 3.5 GeV/$c$.  Photons and electrons are detected in
an array of 8736 CsI(T$\ell$) crystals (ECL)
located inside the magnetic field and covering the entire solid angle of the
charged particle tracking system.  
The 1.5~T magnetic field is returned via an iron
yoke that is instrumented to detect muons and $K_L$ mesons (KLM).
The KLM
consists of alternating layers of resistive plate chambers and
4.7 cm thick steel plates.

%
%

For the light neutral meson $X^0$, 
we use the $\pi^0 \to \gamma \gamma$, 
$\eta \to \gamma\gamma$, $\eta \to \pi^+\pi^-\pi^0$ 
and the $\omega\to\pi^+\pi^-\pi^0$ decay channels.
Charged tracks are required to have impact parameters that are
within $\pm$5 cm of the interaction point along the positron beam axis
and 1 cm in the 
transverse plane.
We reject tracks that are consistent with electrons or muons.
The remaining tracks are identified as pions or kaons according to a 
kaon to pion likelihood ratio. 
Candidate $\pi^0$ mesons are reconstructed from pairs
of photons in the ECL that have an invariant mass  
within $\pm 16$ MeV of the nominal $\pi^0$ mass \cite{pdg}.  
The $\pi^0$ daughter photons are required to have energies greater than
50 MeV. 
Both photons from the $\eta \to \gamma\gamma$ mode are 
required to have $E_{\gamma} > 100$ MeV and
the energy asymmetry of the daughter photons,
$\frac{|E_{\gamma_1}-E_{\gamma_2}|}{E_{\gamma_1}+E_{\gamma_2}}$,
is required to be less than 0.8.
We remove $\eta$ candidates if either of the daughter photons
can be combined with any other photon with 
$E_{\gamma} > 100$ MeV} to form a $\pi^0$ candidate.
Candidate $\eta$ mesons for the $\gamma\gamma$ mode are required to
have 
an invariant mass within $\pm 10.6~{\rm MeV}/c^2$  ($\pm 2.5\sigma$)
of the nominal $\eta$ mass; the corresponding requirement 
for the $\pi^+\pi^-\pi^0$ mode is $\pm 3.4$ MeV/$c^2$.
Candidate $\eta$ mesons are constrained to the nominal $\eta$ mass;
the fit also constrains the $\pi^+\pi^-$ pair from the $3\pi$ channel
to a common vertex point.  
Candidate $\omega$ mesons are $\pi^+\pi^-\pi^0$ 
combinations with an invariant 
mass within $\pm 30$ MeV/$c^2$ of the nominal $\omega$ mass value. 
The CM momentum of the $\pi^0$ from the $\omega$ decay is required to
be greater 
than 350 MeV/$c$ to reduce the large combinatorial background from low
energy photons. 
The $\pi^+ \pi^-$ pair from the $\omega$ decay is required to form
a common vertex point within the beam interaction region taking into
account the lifetime of the $B$ meson. 
Detailed studies of tracking, $\pi^0$ detection, and particle
identification 
yield systematic errors in the detection efficiencies of 7.3\% for
prompt $\pi^0$, 6.6\% for $\eta$, and 7.1\% for $\omega$ mesons.

%
%

For candidate $D^0$ mesons we use the  
$D^0\to K^-\pi^+$, $K^-\pi^+\pi^0$, and
$K^-\pi^+\pi^-\pi^+$ decay modes.
The CM momentum of the $\pi^0$ from $D^0\to K^-\pi^+\pi^0$ decay 
is required to be greater than 300 MeV/$c$. 
The invariant mass of the $D^0$ candidates are required to be within $\pm
2.5\sigma$ of the measured $D^0$ mass, where $\sigma$ is the
$D^0$ mass resolution that varies between 5.5 and 13 MeV$/c^2$
depending on the decay mode. 
A mass and vertex constrained kinematic fit is then performed to the
$D^0$ candidates.  Good $D^0$ candidates are 
required to have an acceptable $\chi^2$ value from the fit.
$D^{*0}$ candidates are reconstructed in the $D^{*0}\to D^0\pi^0$ 
decay mode.  For these $\pi^0$ mesons, the photon energy cut is
reduced to 20 MeV. 
For $D^{*0}$ candidates,
the mass difference, $\Delta m = M(D^0 \pi^0) - M(D^0)$, is 
required to be within $2.5\sigma$ ($\sigma = 0.8 {\rm~MeV}/c^2$) of the
nominal mass difference.  
The systematic error on the $D^0$ meson reconstruction 
efficiency is studied using
the $B^-\to D^0\pi^-$ data sample.  The error is determined to be 15\% 
from a comparison of the observed number of signal events 
relative to the expected yield
assuming the PDG $B^-\to D^0\pi^-$ branching fraction \cite{pdg}. 
The systematic uncertainty in the detection efficiency for low
momentum $\pi^0$s is found to be 10.7\%.

%
%

We combine $D^0$s or $D^{*0}$s with $X^0$ meson candidates to form 
$\bbar$ candidates.  Two kinematic variables are used to identify
signal candidates, the 
beam-constrained mass   
$\mbc = \sqrt{(E_{\rm beam}^{\rm CM})^2 - (p_{B}^{\rm CM})^2}$
and 
the energy difference $\Delta E = E_{B}^{\rm CM} - E_{\rm
beam}^{\rm CM}$, 
where $E_{B}^{\rm CM}$ and $p_{B}^{\rm CM}$ are the CM energy
and momentum of the $\bbar$ candidate, and $E_{\rm beam}^{\rm
CM} = \sqrt{s}/2 \simeq 5.290 {\rm~GeV}$.
The typical $\mbc$ resolution is 3 MeV$/c^2$; the
$\Delta E$ resolution ranges from 17 to 25~MeV,
depending on the decay mode.
When more than one $\bbar$ candidate is found in an event,
the candidate with the minimum $\chi^2$ is chosen, where 
$\chi^2 = \chi^2_{D^0} + \chi^2_{X^0} (+ \chi^2_{\Delta m}$).
Here $\chi^2_{D^0}$ is the $\chi^2$ of the kinematic
fit to the $D^0$, $\chi^2_{X^0}$ is the $\chi^2$ of the kinematic
fit to either the $\pi^0$ or the $\eta$.  For the $\omega$, 
$\chi^2_{X^0} = (\Delta(M_{\omega})/\sigma(M_{\omega}))^2$, 
where $\sigma$ is the measured resolution.  
For the $D^{*0}X^0$ modes, 
$\chi^2_{\Delta m}$, defined as $(\Delta(\Delta m)/\sigma({\Delta m}))^2$,
is included in the best candidate selection.

%
%

The background from continuum $e^+ e^- \to q\overline{q}$
production is suppressed in the following ways.  
For the $D^{(*)0}\eta$ final state, we apply cuts 
on the ratio of the second to  
zero-th Fox-Wolfram moments\cite{fw}, $R_2$, 
and the angle between the thrust axis~\cite{thrust} of the $B$ 
candidate and the thrust axis of the rest of the event 
($\cos{\theta_{T}}$).
For the $D^{(*)0}\pi^0$ and $D^{(*)0}\omega$ final states, 
we use the $B$ flight direction and a Fisher discriminant\cite{fisher} 
containing several variables that quantify event topology\cite{sfw}. 
We also use the $D^{*0}$ helicity angle for the $D^{*0} \pi^0$ mode
and the $\omega$ helicity angle and the $\omega$ decay amplitude for
the $D^{(*)0} \omega$ mode \cite{helicity}.  Each of these variables
is 
parameterized to form signal (S) and background (BG) probability
density functions (PDF).  The PDFs are multiplied to form a single
likelihood ${\cal L}_{\rm S (BG)}$, and then a cut is applied on
the likelihood ratio ${\cal L}_{\rm S}/({\cal L}_{\rm S}+{\cal L}_{\rm
BG})$ to suppress the $q\overline{q}$ background.   
Signal PDFs are determined using Monte Carlo (MC) and background PDFs
are obtained 
from $\mbc$ sideband data. The cut efficiencies are typically 70\% 
and remove more than 90\% of the $q\overline{q}$ background.  The 
systematic error in the efficiency for this cut is determined by
applying the same 
procedure to the $B^- \to D^{*0}\pi^-$ data sample.  By comparing the 
cut efficiency between the data sample and MC, the systematic error 
is determined to be 5\%.

%
%

In addition to the $q\overline{q}$ background, we observe large
background contributions from color favored $B \to D^{(*)} (n\pi)^-$
decays and cross-talk from $D^{*0} X^0$ to $D^0 X^0$ modes.  
The $D^{*+}\rho^-$ mode has the same final state as $D^0\omega$ and
$D^0 \eta$.  
However, when the $D^{*+}$ daughter pion is combined with the
$\rho^-$, the invariant mass rarely falls within the $\omega$ or
$\eta$ mass window.  The $D^{(*)0}\rho^-$ final state contaminates
the $D^{(*)0}\pi^0$ mode if the $\rho^-$ decays to a fast $\pi^0$.
This mode also contaminates the $D^{(*)0}\eta$ channel if a photon
from the fast $\pi^0$ is combined with another photon to form an $\eta$
candidate.  
About half of these events are removed by explicitly reconstructing
the $D^{(*)0}\rho^-$ final state.  
The contributions of these backgrounds in the $\eta$ channel, as
well as the feed-across 
from the $D^{(*)0}\pi^0$ mode, is also minimized by the $\pi^0$ veto
discussed above.  
We also check for background contributions from
$\overline{B}\to D^{(*)0}\rho^{\prime -}$ ($\rho^{\prime -}\to
\omega\pi^-$) decays which have recently 
been observed by CLEO\cite{cleo0103021}.
This two-body decay produces 
high momentum $D^{(*)0}$s and $\omega$s that can fake signal events.
Monte Carlo studies indicate that the remaining background events are
shifted in $\Delta E$ by approximately the mass of the missing slow
pion and thus can be distinguished from signal events by fitting the
$\Delta E$ distribution.

%
%

\begin{figure}
\begin{center}
\vspace{-36pt}
\epsfig{file=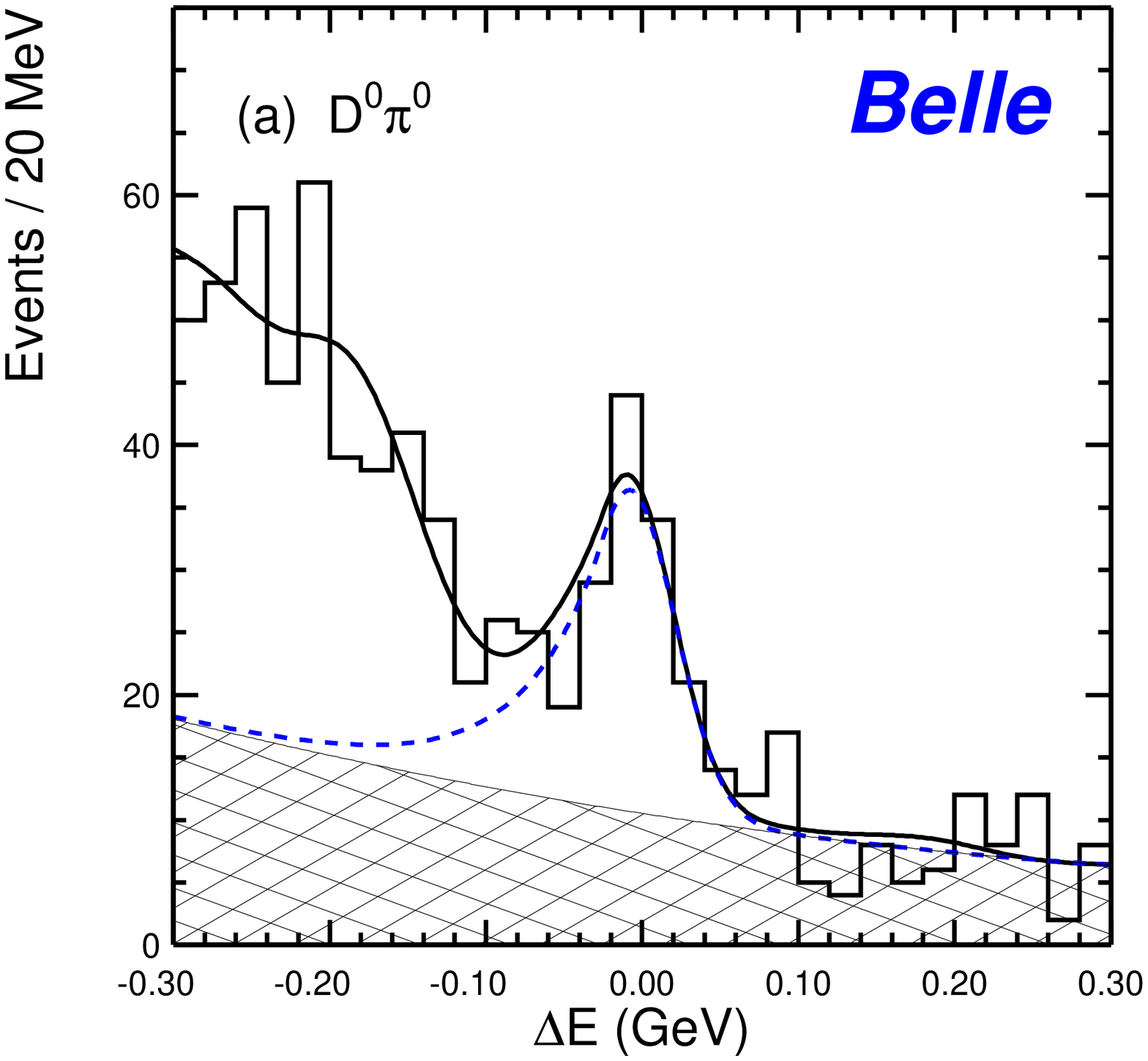, width=3in}
\epsfig{file=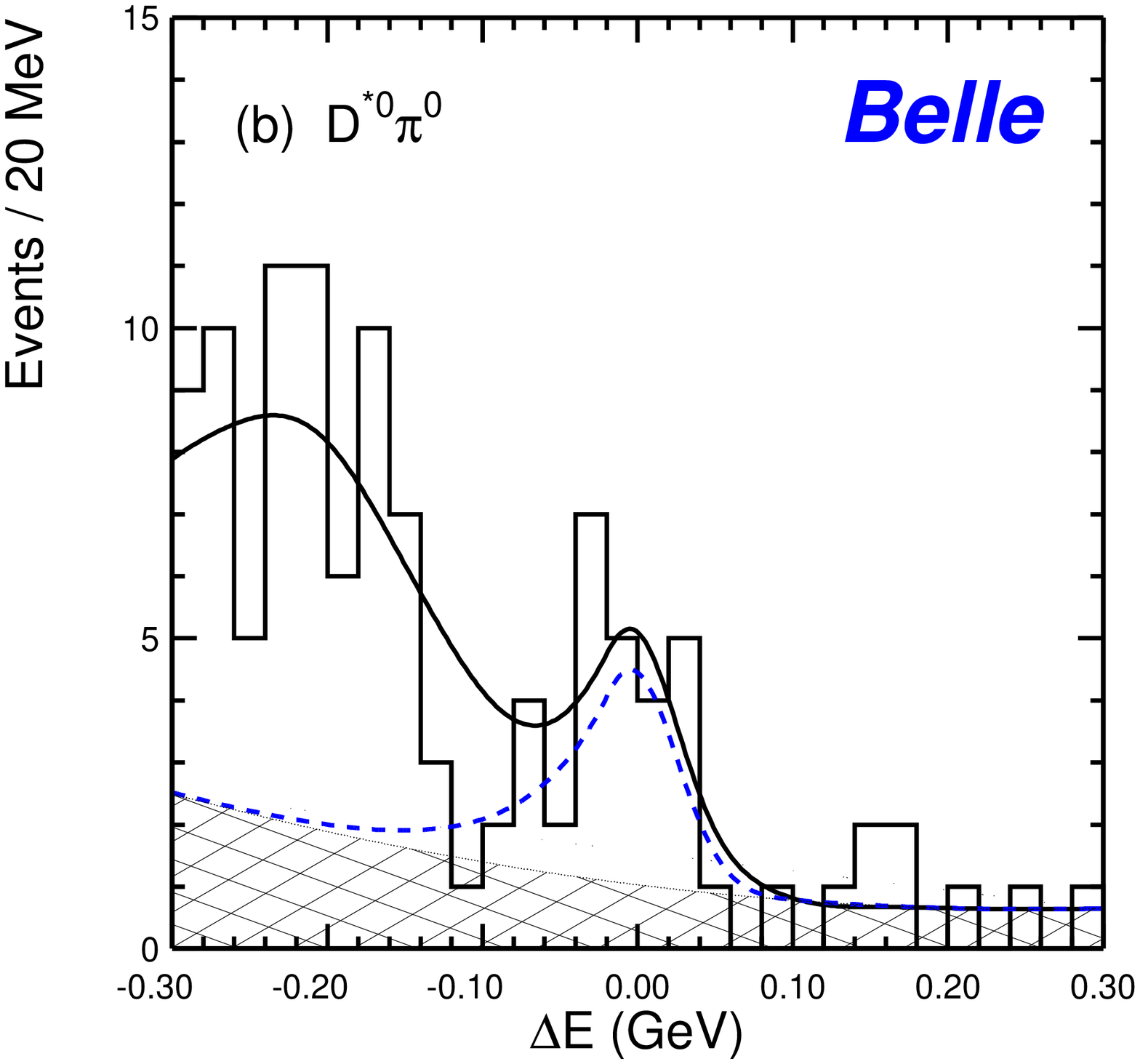, width=3in} \\
\vspace{-36pt}
\epsfig{file=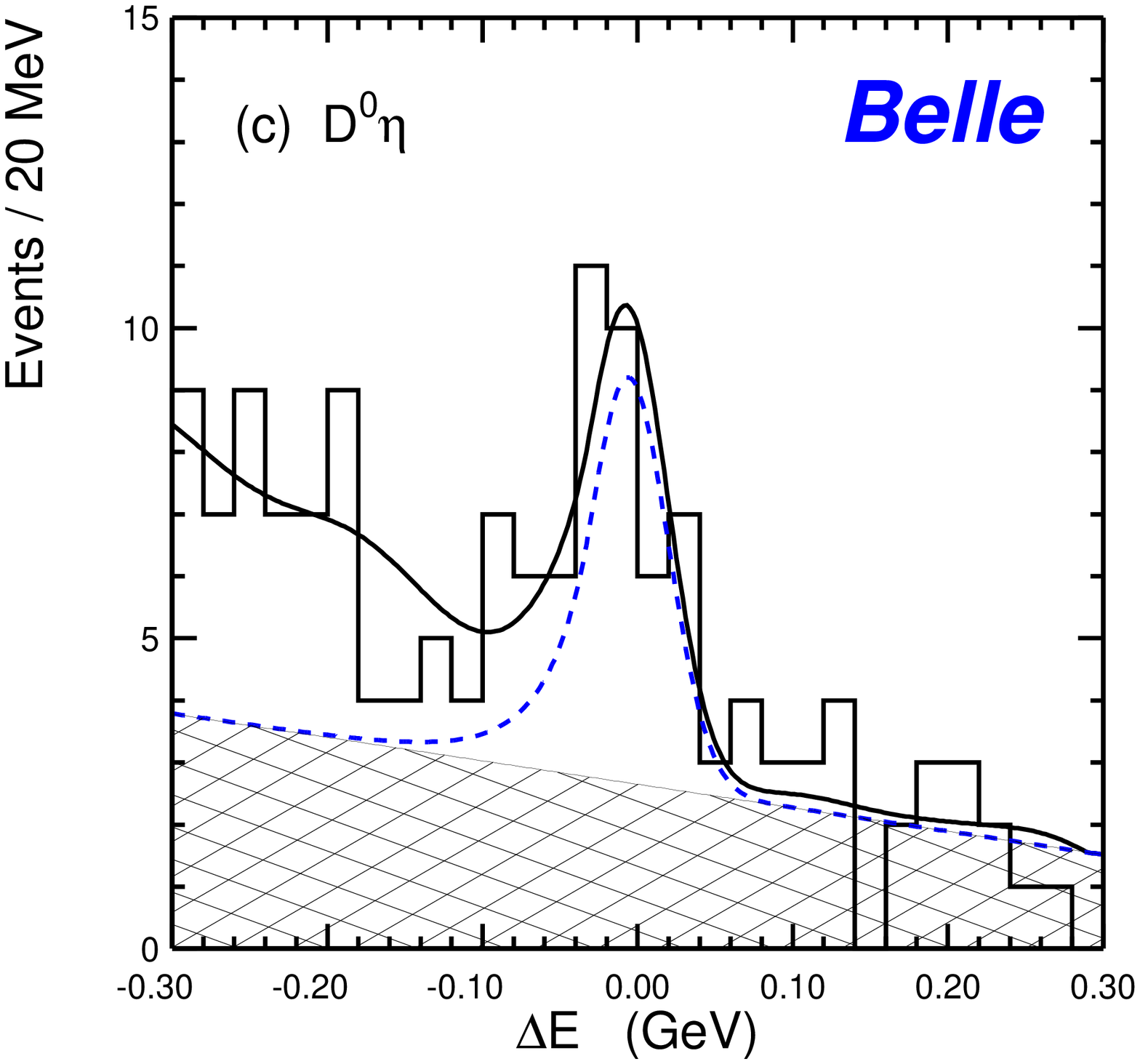, width=3in}
\epsfig{file=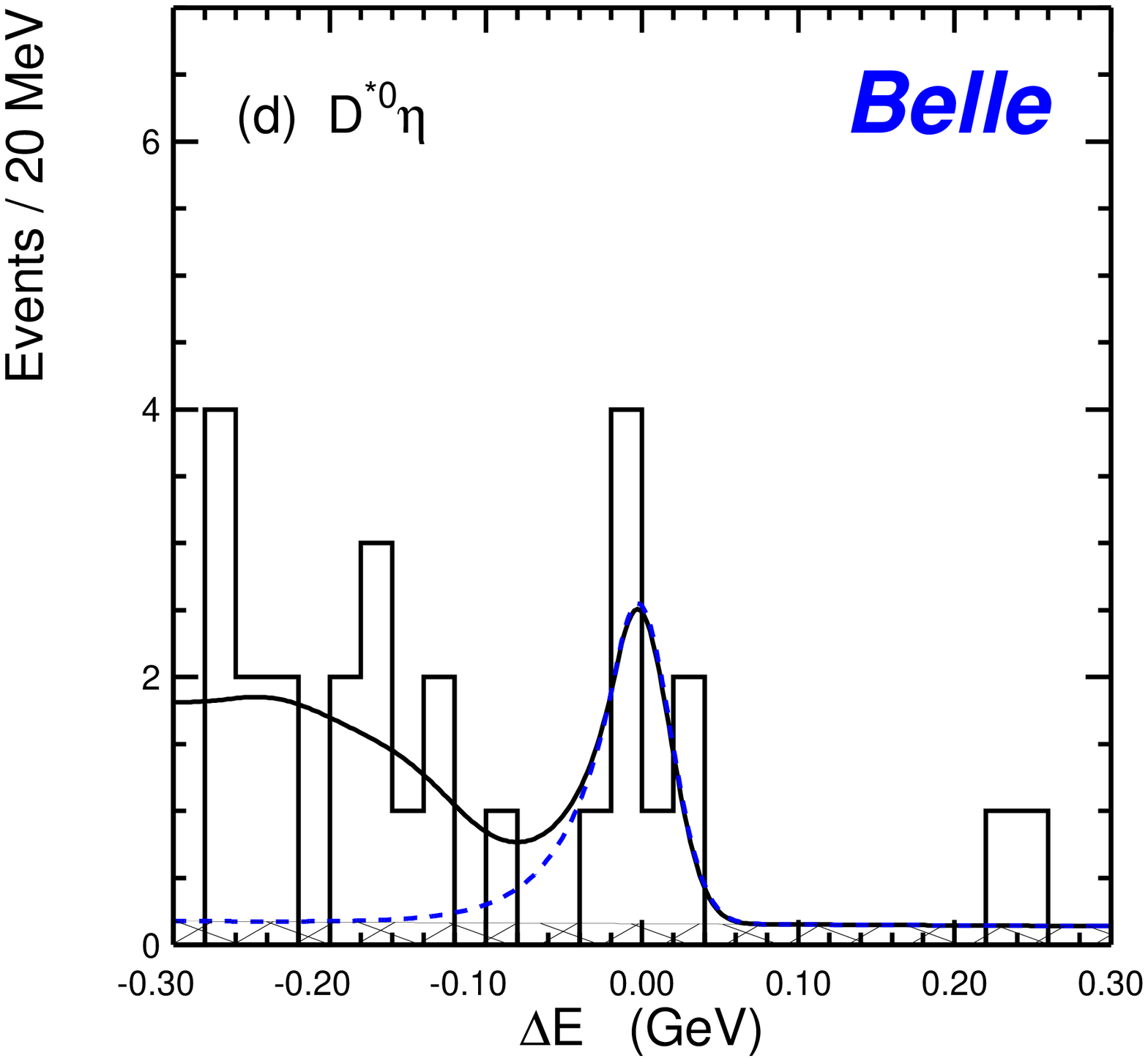, width=3in} \\
\vspace{-36pt}
\epsfig{file=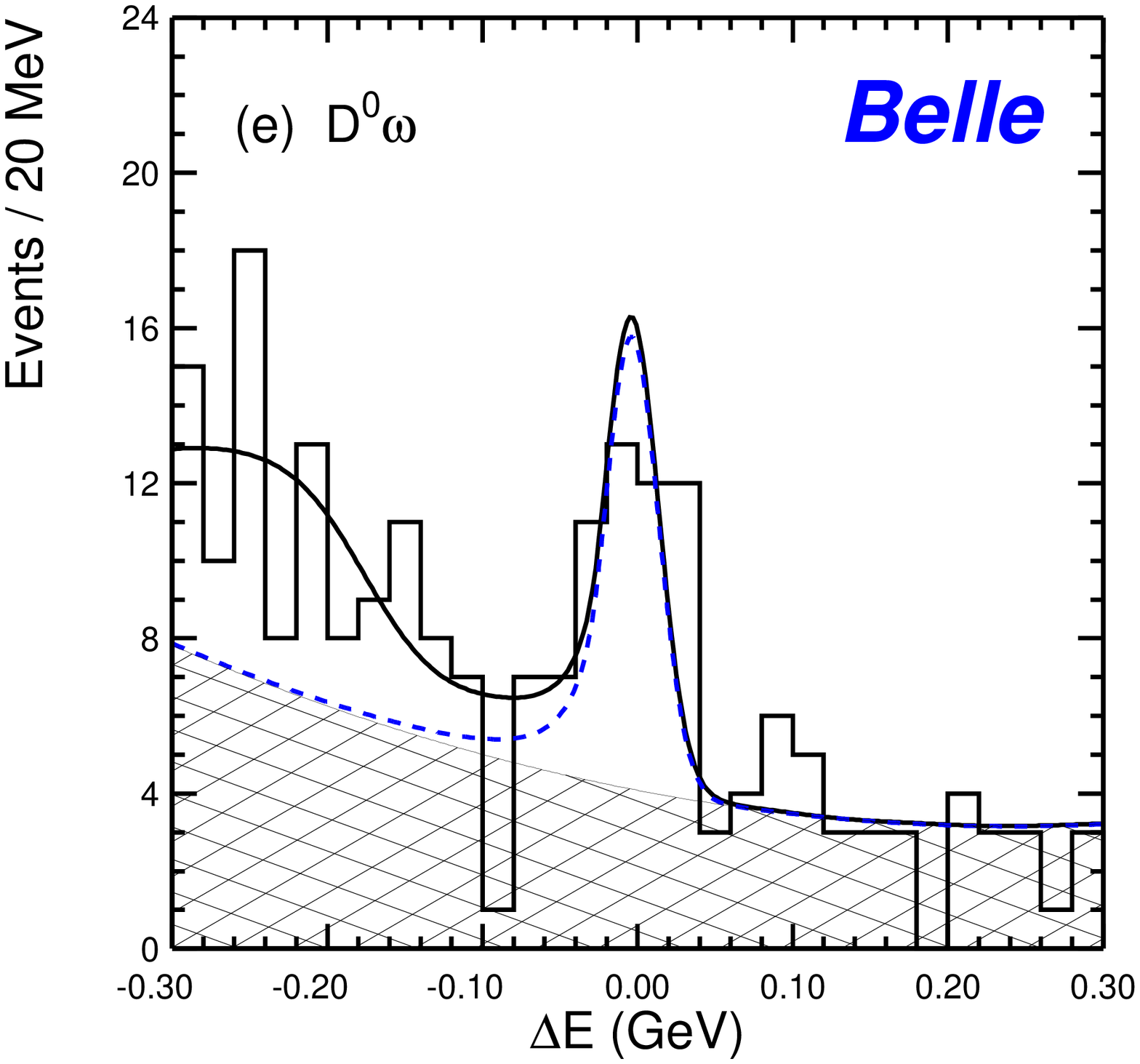, width=3in}
\epsfig{file=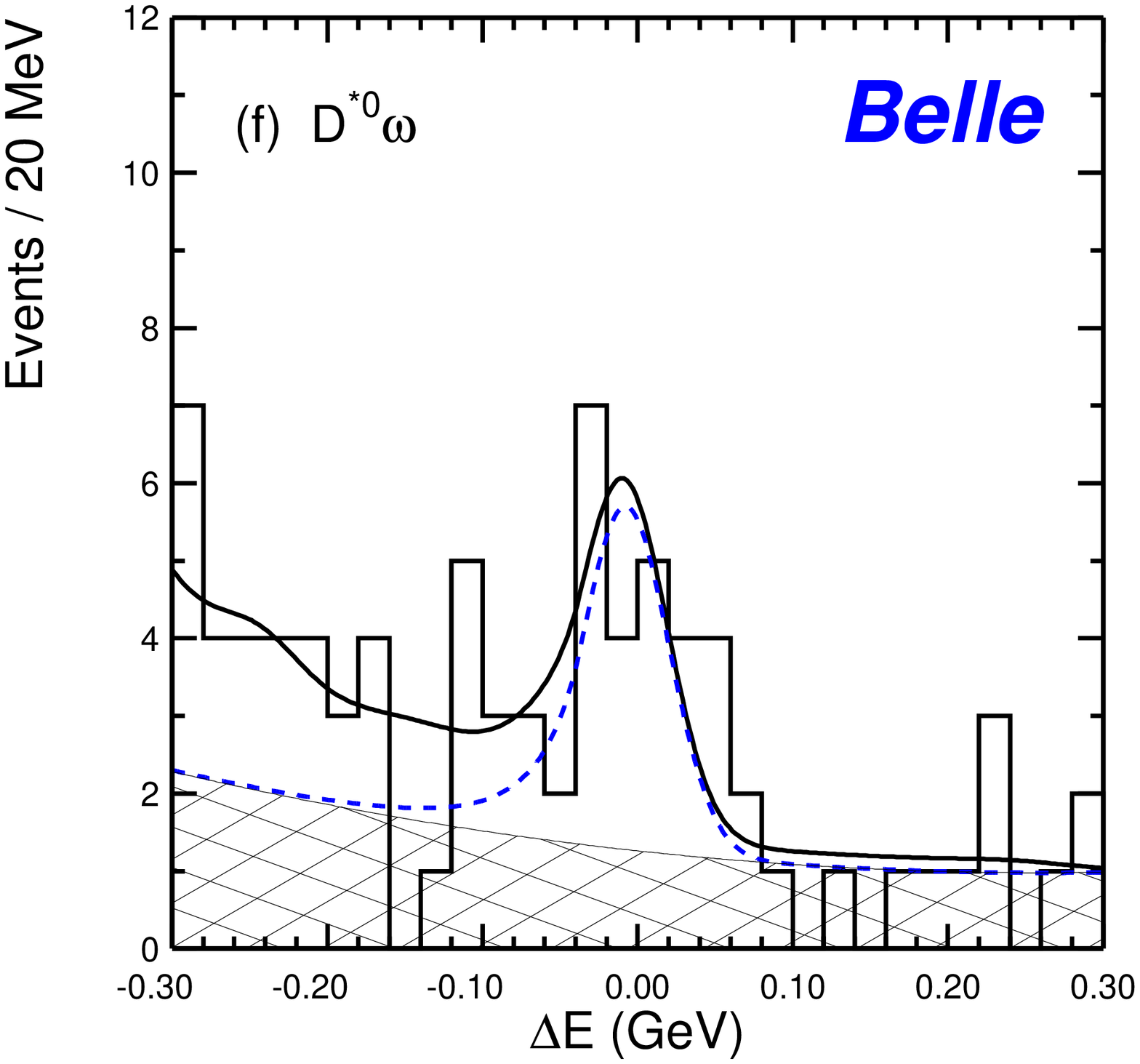, width=3in}
\medskip
\caption{The $\Delta E$ distribution for (a) $D^{0}\pi^0$, 
(b) $D^{*0}\pi^0$, (c) $D^{0}\eta$, 
(d) $D^{*0}\eta$, (e) $D^{0}\omega$, and 
(f) $D^{*0}\omega$.
The solid line shows the fitting result.  The dashed line shows the
sum of the signal component and the combinatorial background component.  
The combinatorial component is shown separately as the cross-hatched area.}
\label{fig-dpi0}
\end{center}
\end{figure}

The $\Delta E$ distributions for the various $D^{(*)0}X^0$ decays are
shown in Fig.~\ref{fig-dpi0} after applying all selection 
cuts and requiring $\mbc$ to be between 5.272 GeV/$c^2$ and 5.288
GeV/$c^2$.
The signal is modeled with a Crystal-Ball function\cite{cbline} with
parameters obtained from MC.  
The background functions include a combinatorial component and a
color-favored component.  The $D^0 X^0$ modes also include a component
for cross-talk from color-suppressed $D^{*0} X^0$ modes.  
The combinatorial component is taken to be a second order polynomial 
with parameters determined by the $\Delta E$ shape in the $\mbc$
sideband (5.20 GeV/$c^2 < \mbc < 5.26$ GeV/$c^2$). The shapes of the
color-favored and cross-talk components are modeled by MC 
histograms.  The signal and background normalizations are free
parameters in each fit.

Table~\ref{yields-table} lists the signal yield, statistical
significance, 
efficiency, and the branching fraction for each $D^{(*)0}X^0$ mode. 
The systematic errors due to fitting are obtained 
by varying the parameters of the fitting functions within $1\sigma$ of
their nominal values. The change in the signal yield from each variation is
added in quadrature to obtain the fitting systematic errors.  
These are typically 10\%. 
The statistical significance is defined as 
$\sqrt{-2{\rm ln}({\cal L}(0)/{\cal L}_{\rm max})}$
where ${\cal L}_{\rm max}$ is the likelihood at the nominal signal
yield and ${\cal L}(0)$ is the likelihood with the signal yield fixed
to zero.  We observe signals for $\bbar \to D^0 \pi^0$, $D^0 \omega$, 
and $D^{*0} \omega$ decays with more than $4\sigma$ significance.  We
find evidence of signals for $\bbar \to D^{*0}\pi^0$, 
$D^0 \eta$, and $D^{*0} \eta$ with more than $3\sigma$ significance.  
For decay modes with significance less than 4, we give 90\%
confidence level upper limits (UL) on the signal yields ($N_{\rm
S}^{\rm UL}$) from the relation 
$\int_0^{N_{\rm S}^{\rm UL}}{\cal L}(N_{\rm S}) \, dN_{\rm S} \,/\, 
\int_0^{\infty}{\cal L}(N_{\rm S}) \, dN_{\rm S} =0.9$, 
where ${\cal L}(N_{\rm S})$ denotes the maximum
likelihood with the signal yield fixed at $N_{\rm S}$. 
The efficiencies for each decay mode are calibrated with control data 
samples. 
The final systematic errors include the errors in fitting,
reconstruction efficiency, background suppression cut 
efficiency, and the number of $B\overline{B}$ pairs. 
Assuming 
the number of $B^0\overline{B}{}^0$ and $B^+B^-$ pairs are equal, we
calculate the branching fractions for various decay modes given in 
Table~\ref{yields-table}.  
The branching fraction upper limits are calculated by increasing
$N_{\rm S}^{\rm UL}$ and reducing the efficiency by their systematic errors.

%
%

In summary, using 22.8 million $B\overline{B}$ events collected with the
Belle detector, we report the first observations of color-suppressed 
$\bbar\to D^0 \pi^0$ and $D^{(*)0}\omega$ decays.  We also find 
evidence for $\bbar \to D^{*0} \pi^0$ and $D^{(*)0} \eta$ signals. 
All the color-suppressed modes have similar branching fractions with
central values between 1.4 and 3.4 $\times 10^{-4}$, as shown in
Table~\ref{yields-table}.
In general, the branching fractions are consistently higher than 
recent theory predictions\cite{neubert} based on 
the factorization hypothesis.  
This may be accounted for by additional corrections to the
factorization models, or by non-factorizable effects such as final
state interactions.

%
%


We wish to thank the KEKB accelerator group for the excellent
operation of the KEKB accelerator. We acknowledge support from the
Ministry of Education, Culture, Sports, Science, and Technology of Japan
and the Japan Society for the Promotion of Science; the Australian
Research
Council and the Australian Department of Industry, Science and
Resources; the Department of Science and Technology of India; the BK21
program of the Ministry of Education of Korea and the CHEP SRC
program of the Korea Science and Engineering Foundation; the Polish
State Committee for Scientific Research under contract No.2P03B 17017;
the Ministry of Science and Technology of Russian Federation; the
National Science Council and the Ministry of Education of Taiwan; the
Japan-Taiwan Cooperative Program of the Interchange Association; and
the U.S. Department of Energy.

%
%

\newpage

%
%

\begin{table}[htb]
\caption{The obtained signal yield, statistical significance,
efficiency including the sub-decay branching fractions,  branching
fraction ($\mathcal{B}$), and $90\%$ confidence level upper 
limit (UL) for each $\bbar\to D^{(*)0}X^0$ decay mode.}
\label{yields-table}
\medskip
\begin{tabular}{llcclc}
Mode & Signal Yield & Significance & Efficiency(\%) & $\mathcal{B}$
($\times 10^{-4}$) & UL ($\times 10^{-4}$) \\ \hline
$D^0\pi^0$      & $127.6\;^{+18.5\; +11.6}_{-17.9\; -12.5}$ & 7.9 & 1.93 & 
$2.9\;^{+0.4}_{-0.3} \pm 0.6 $  & -- \\
$D^{*0}\pi^0$   & $17.1\;^{+6.6\; +1.6}_{-5.9\; -2.4}$ & 3.2 & 0.49 & 
$1.5\;^{+0.6\; +0.3}_{-0.5\; -0.4} $  & 2.3 \\  \hline
$D^0\eta$       & $ 25.7\;^{+8.4}_{-7.7}{}\;^{+3.0}_{-2.8}$ 
        & 3.8 & 0.79 & $ 1.4\;^{+0.5}_{-0.4} \pm 0.2 $ & 2.1 \\
$D^{*0}\eta$    & $ 7.7\;^{+3.4}_{-2.7} {}\;^{+0.7}_{-0.8} $ 
        & 3.6 & 0.22 & $ 1.5\;^{+0.7}_{-0.6} \pm 0.4 $ & 2.7 \\ \hline
$D^0\omega$     & $30.2\;^{+8.6\; +3.1}_{-7.8\; -3.4}$ & 4.7 & 0.80 & 
$1.7\;^{+0.5\; +0.3}_{-0.4\; -0.4} $ & -- \\
$D^{*0}\omega$  & $17.7\;^{+6.5\; +2.3}_{-5.8\; -2.2}$ & 4.3 & 0.23 & $
3.4\;^{+1.3}_{-1.1}\pm 0.8 $ & -- \\ 
\end{tabular}
\end{table}

%
%

\vspace{2cm}

\clearpage

%
%

\begin{thebibliography}{99}
\bibitem{sfw}{
We create a seven variable Fisher containing $\cos{\theta_{T}}$
and $S_\perp$, 
a variable that measures the momentum flow transverse to the $X^0$
direction.  The next two variables, motivated by the original
Fox-Wolfram moments, are defined as 
$r_l =
(\Sigma_{i,j}{|p_i||p_j|P_l(\cos{\theta_{i,j}})})/(\Sigma_{i,j}{|p_i||p_j|})
$ 
where $i$ corresponds to a $B$ candidate daughter particle and 
$j$ corresponds to a particle not associated with the $B$ candidate; 
$p$ is the particle's momentum, $P_l$ is a Legendre polynomial, 
$\theta_{i,j}$ is the angle between particles $i$ and $j$, and $l$ is
either $2$ or $4$.  
The remaining three variables are equivalent in form to $r_l$ 
with the exception that neither particles $i$ or $j$ are associated with the 
$B$ candidate, $i \ne j$, and $l$ corresponds to $2$, $3$, and $4$.
}
\bibitem{helicity}{
The $D^{*0}$ helicity angle is defined as the angle between the 
$D^{*0}$ flight direction and the direction of the $D^{*0}$'s daughter
in the $D^{*0}$ rest frame. 
The $\omega$ helicity angle is defined as the angle between 
the $B$ flight direction and the normal to the $\omega$ 
decay plane in the $\omega$ rest frame.  The $\omega$ decay amplitude
is defined as the cross product of the two charged pions in the
$\omega$ rest frame. 
}
\end{thebibliography}
\end{document}